\documentclass[conference,a4paper]{IEEEtran}

\usepackage{amsmath}
\usepackage[pdftex]{graphicx}
\usepackage{url}
\usepackage{kotex}

\ifCLASSOPTIONcompsoc
  \usepackage[caption=false,font=normalsize,labelfont=sf,textfont=sf]{subfig}
\else
  \usepackage[caption=false,font=footnotesize]{subfig}
\fi

\makeatletter

\@addtoreset{figure}{section}
\makeatother

\begin{document}
\title{When Do Users Change Their Profile Information on Twitter?}

\author{
    \IEEEauthorblockN{Jinsei Shima \quad Mitsuo Yoshida \quad Kyoji Umemura}
    \IEEEauthorblockA{Department of Computer Science and Engeneering,\\
    Toyohashi University of Technology,\\
    Aichi, Japan\\
    j163336@edu.tut.ac.jp, yoshida@cs.tut.ac.jp, umemura@tut.jp}
}

\maketitle

\begin{abstract}
We can see profile information such as name, description and location in order to know the user on social media.
However, this profile information is not always fixed.
If there is a change in the user's life, the profile information will be changed.
In this study, we focus on user's profile information changes and analyze the timing and reasons for these changes on Twitter.
The results indicate that the peak of profile information change occurs in April among Japanese users, but there was no such trend observed for English users throughout the year.
Our analysis also shows that English users most frequently change their names on their birthdays, while Japanese users change their names as their Twitter engagement and activities decrease over time.
\end{abstract}

\renewcommand\IEEEkeywordsname{Keywords}

\IEEEpeerreviewmaketitle

\section{Introduction}

When is the last time you changed your profile information such as name, description and location on social media?
If you change jobs, you will add a new job title to the description on your profile.
If you move, you will update your location.
If you are separated from your partner, you will change and indicate that you are now looking for a new partner.
As we can see, social media users change their profile information for various reasons.
Changes in their lives will affect their profile information.
We are interested in discovering the regularity of profile information changes.

Twitter\footnote{\url{https://twitter.com/}} is one of the most popular social media which has been frequently used by many global users in recent years.
The profile is used to better understand the user's background on Twitter, and the user can freely change it at any time.
Twitter is used all over the world, and the timing of changing profile information may vary depending on the language sphere.
Also, when changing the profile, there is a possibility that keywords related to changes in specific events and living environments are inserted.

Many researchers have been attempting to analyze and utilize the profile information on Twitter.
Semertzidis et al.~\cite{Semertzidis2013} analyzed the profile information in order to understand what Twitter users choose to expose about themselves in their profile information.
Mislove et al.~\cite{Mislove2011} used the profile information in order to compare the Twitter population to the U.S. population along three axes (geography, gender, and race/ethnicity).
Alowibdi et al.~\cite{Alowibdi2013} and Vicente et al.~\cite{Vicente2015} used the profile information to conduct a gender classification on Twitter.
Uddin et al.~\cite{Uddin2014} used the profile information to understand types of users such as humans, bots, spammers, businesses and professionals.
Vainio and Holmberg~\cite{Vainio2017} used the profile information to identify the user's research field.

Previous study has assumed that the user's profile information remains unchanged.
However in reality, users often change their profile information.
We focus on user's profile information changes and analyze the timing and reasons for change on Twitter.
In this paper, we address the following two research questions:
\begin{description}
 \item[RQ1]When do users change their profile information on Twitter?
 \item[RQ2]What keyword should be inserted when the profile information is changed?
\end{description}
The results indicate that the peak of profile information change occurs in April among Japanese users, but there was no such trend observed for English users throughout the year.
Our analysis also shows that English users most frequently change their names on their birthdays, while Japanese users change their names as their Twitter engagement and activities decrease over time.

\section{Data}

In this study, we use Japanese (ja) and English (en) retweets on Twitter collected from 2015 to 2016.
The details of these data are shown in TABLE~\ref{tb:retweet-count}.
These data were collected using the Twitter Search API\footnote{\url{https://developer.twitter.com/en/docs/tweets/search/api-reference/get-search-tweets}}.
Normally, tweets collected using the Twitter Sample Tweets API\footnote{\url{https://developer.twitter.com/en/docs/tweets/sample-realtime/api-reference/get-statuses-sample}} are used in such study.
Tweets we can collect with this Sample Tweets API are less than 1\% on Twitter.
Thus, we used retweets collected using the Search API\footnote{We constantly searched by query ``RT lang:ja'' and ``RT lang:en''.} because we need more tweets.
Since many users are using retweets~\cite{Liu2014}, we assume that sampled ``retweeting'' Twitter users are same as usual Twitter users.

For example, users registered in June 2016 cannot change their profile information in March 2016.
We select users in 2015 data and monitor their profile information in 2016 data.
Let $ U_{2015} $ be a set of users who posted in 2015 and Let $ U_{2016} $ be a set of users who posted in 2016.
The set we use for analysis is $ U_{2015} \cap U_{2016} $.
The numbers of users are shown in TABLE~\ref{tb:user-count}.

\begin{table}[tp]
  \centering
  \caption{The number of Retweet and Users Collected}
  \begin{tabular}{ c c r r }
\hline
lang & year & \# of retweets & \# of users \\
\hline \hline
ja & 2015 & 9,441,746,992 & 24,166,859 \\
ja & 2016 & 7,605,744,928 & 28,432,609 \\
\hline
en & 2015 & 16,170,390,938 & 93,618,760 \\
en & 2016 & 14,058,551,334 & 93,903,809 \\
\hline
  \end{tabular}
  \label{tb:retweet-count}
\end{table}
    
\begin{table}[tp]
  \centering
  \caption{The number of users used for analysis.}
  \begin{tabular}{ c r }
\hline
lang  & \# of users \\
\hline \hline
ja & 8,083,170 \\
en & 19,485,950 \\
\hline
   \end{tabular}
  \label{tb:user-count}
\end{table}

\begin{figure}[tp]
  \centering
  \subfloat[Japanese (ja)]{
    \includegraphics[width=1\linewidth]{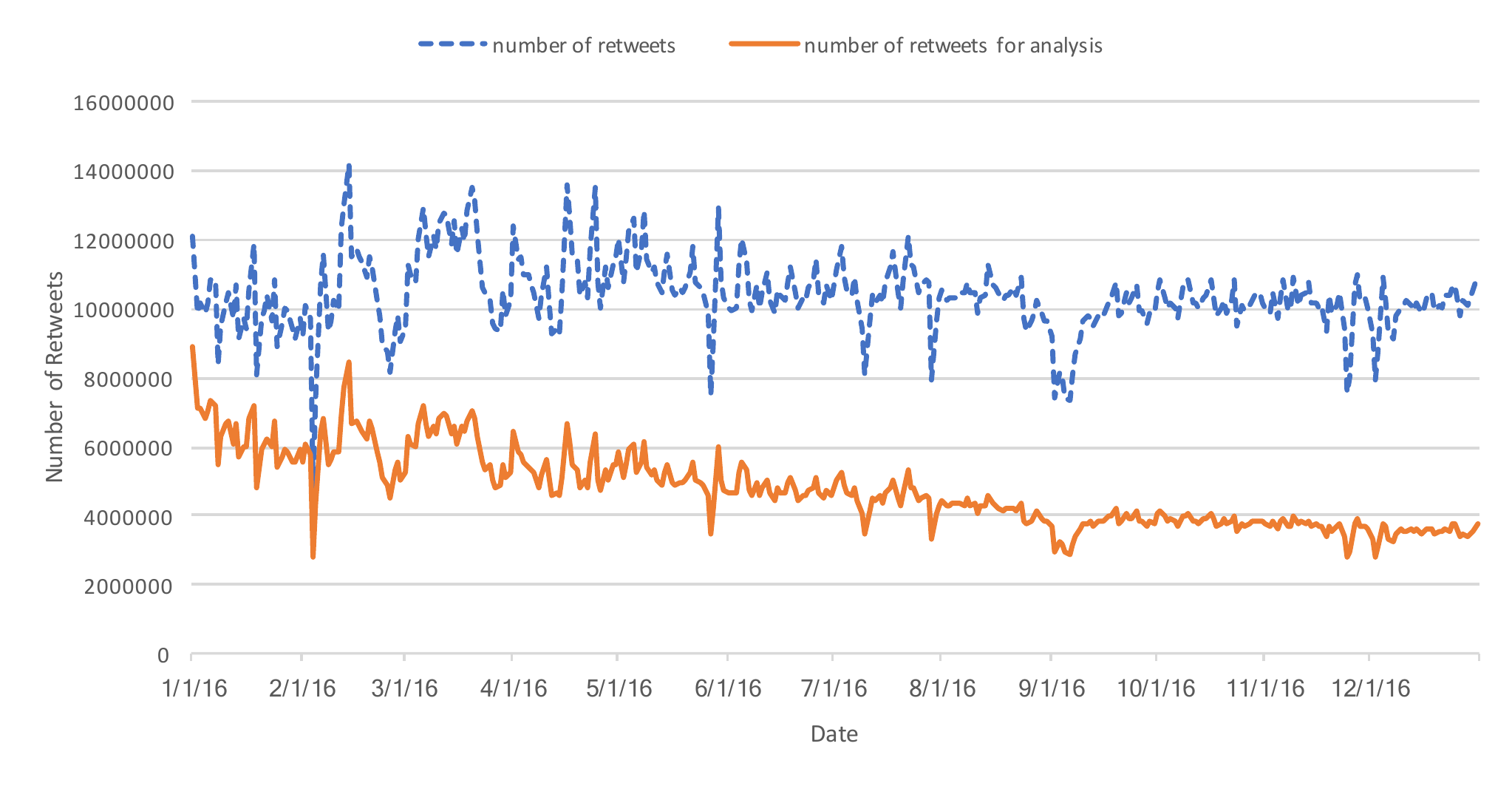}
    \label{fig:retweet-count-ja}
  }
  \hfil
  \subfloat[English (en)]{
    \includegraphics[width=1\linewidth]{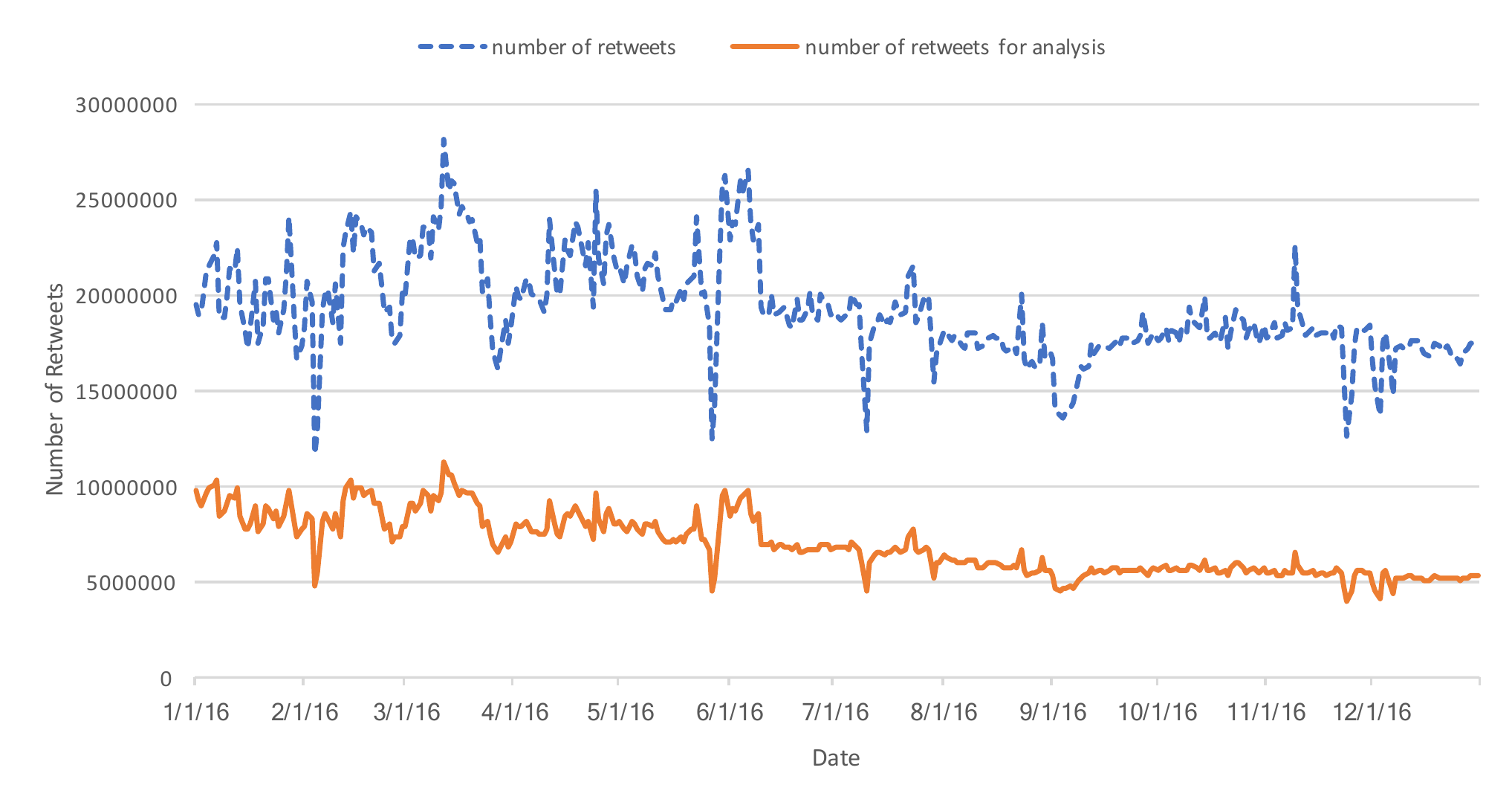}
    \label{fig:retweet-count-en}
  }
  \caption{The number of retweets collected per day: data collection amount dropped on some days. The solid line is the number of retweets limited to users who posted in 2015. The user activity is gradually decreasing.}
  \label{fig:retweet-count}
\end{figure}

The number of retweets collected per day is indicated by the dashed lines in Fig.~\ref{fig:retweet-count}.
The data collection amount dropped on some days, for example February 4, May 27 and September 6.
This happens when Twitter APIs and our collecting server are faulty or experiencing other technical issues.
The solid line is the number of retweets limited to users who posted in 2015.
The activity of the user is gradually decreasing.
The above situations may affect the results of the present study.

For analysis, we use only the retweet date (``created\_at'') and the user object\footnote{\url{https://developer.twitter.com/en/docs/tweets/data-dictionary/overview/user-object}} among tweet objects\footnote{\url{https://developer.twitter.com/en/docs/tweets/data-dictionary/overview/tweet-object}}.
In addition, we focus on ``name'', ``description'' and ``location'' as user's profile information, unlike previous studies~\cite{Liu2014,Jain2016} focused on ``screen\_name''.
A real name or a nickname may be written in the name field.
It may include a current status (e.g., holiday and busy).
A brief self-introduction may be written in the description field, and an actual residential area may be written in the location field.
In our analysis, the retweet date is converted into Japan Standard Time (JST) in Japanese and English retweets.
For example, in the analysis on the United States, the date will shift due to the time difference.

\section{Results and Analysis}

\subsection{RQ1: When to Change User's Profile}

In this section, we investigate the trend of the date users change their profile information.
In Japan, new lives (e.g., university admission and employment) often start from April.
If life changes affect profile information, we will observe many profile information changes occurring in April.

\begin{figure}[tp]
  \centering
  \subfloat[Japanese (ja)]{
    \includegraphics[width=1\linewidth]{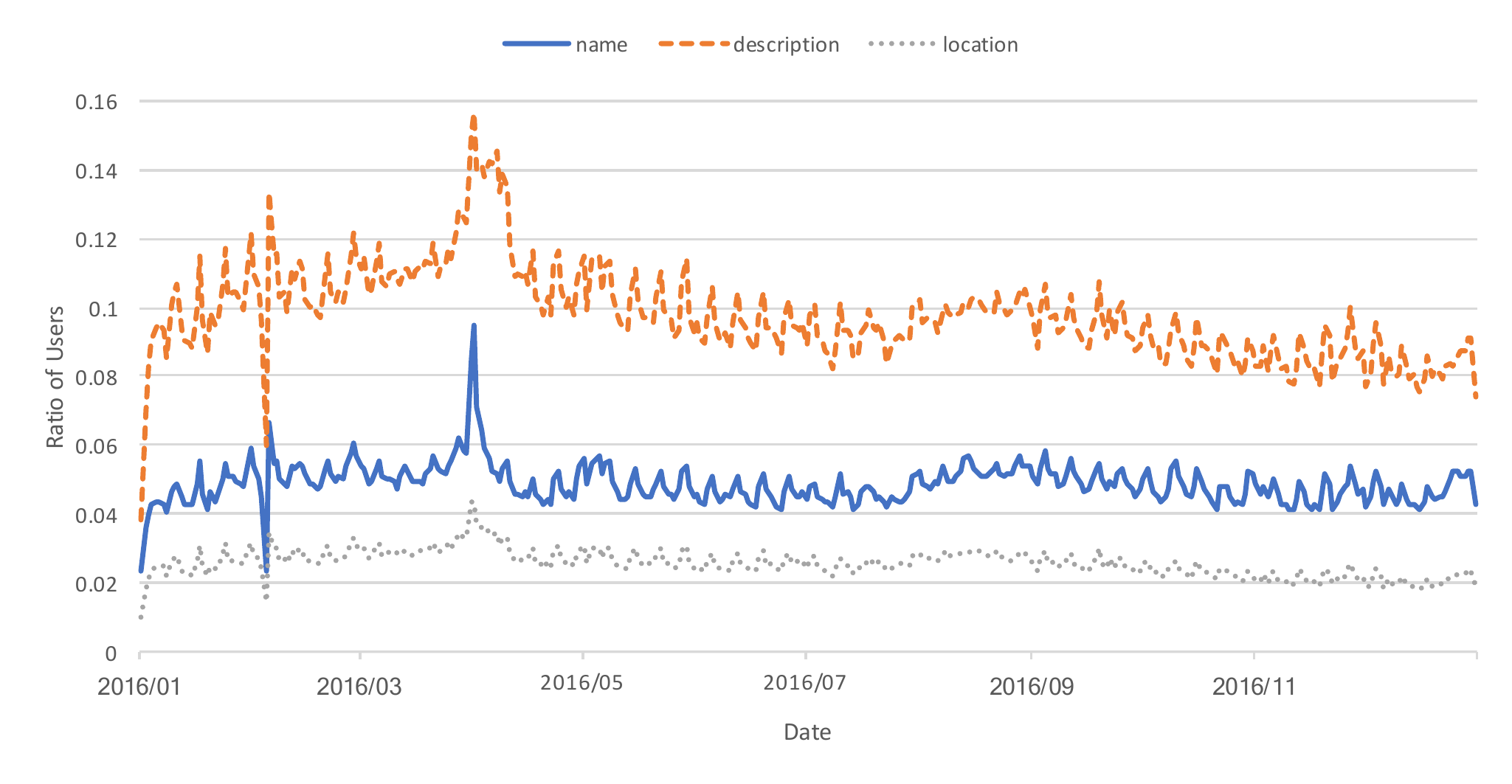}
    \label{fig:change-profile-ratio-ja}
  }
  \hfil
  \subfloat[English (en)]{
    \includegraphics[width=1\linewidth]{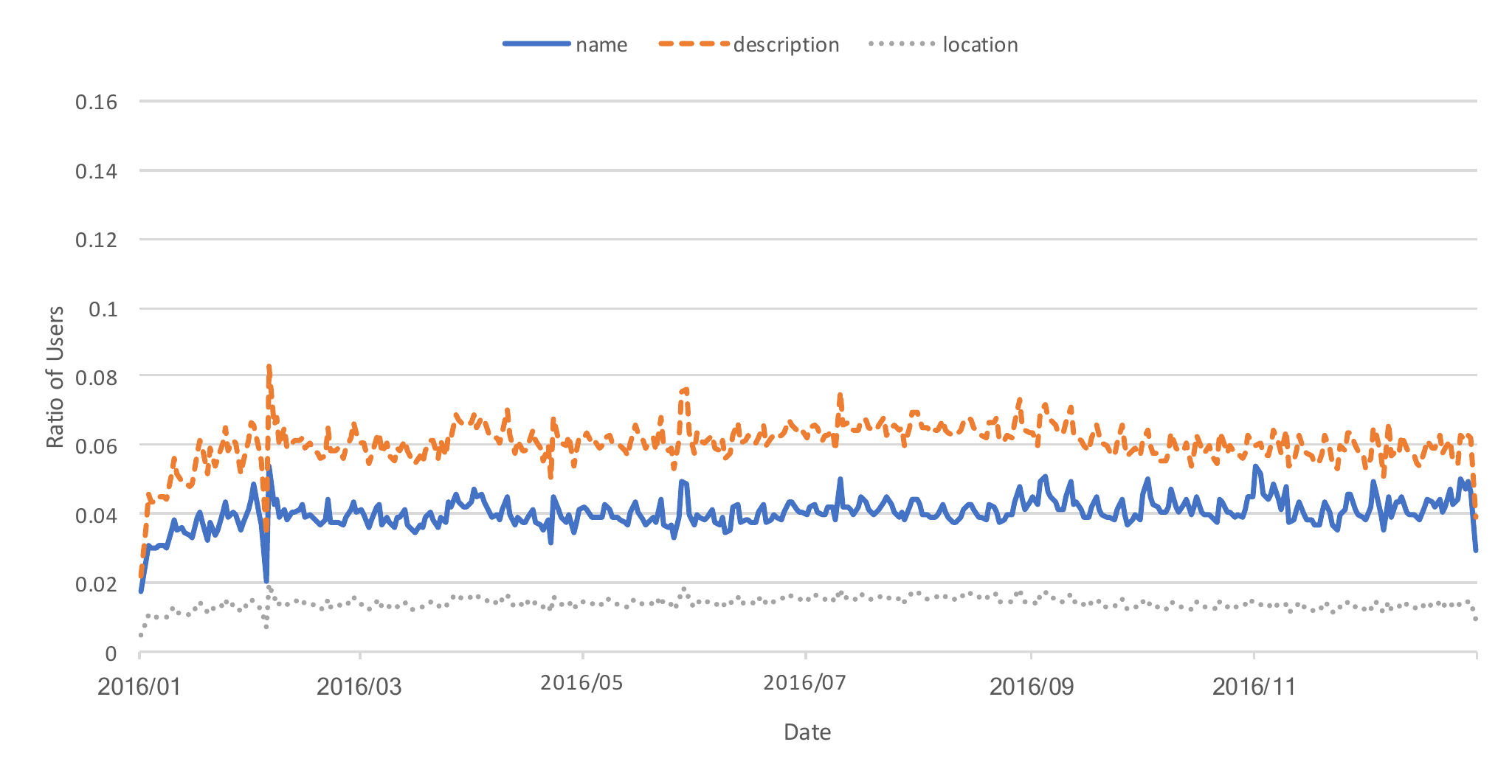}
    \label{fig:change-profile-ratio-en}
  }
  \caption{Trend of profile information change: the peak of profile information change occurs in April in Japanese, but there was no change trend in English throughout the year.}
  \label{fig:change-profile-ratio}
\end{figure}

In Fig.~\ref{fig:change-profile-ratio}, we show the trends of profile information change dates in Japanese and English.
The vertical axis indicates the ratio of users who changed their profile information on that day.
The denominator of the ratio is the number of users retweeting on that day.

The description was the most changeable in both Japanese and English.
As far as observing the results, Japanese users more frequently change their profile information over English users.
We found that a peak occurs on April 1 in Japanese.
This may be the effect of the new life or April Fools' Day.
There was no change trend in English as much as Japanese, but a small peak of name change was found in November.
This is consistent with the presidential election in the United States.

\begin{figure}[tp]
  \centering
  \subfloat[Japanese (ja)]{
    \includegraphics[width=1\linewidth]{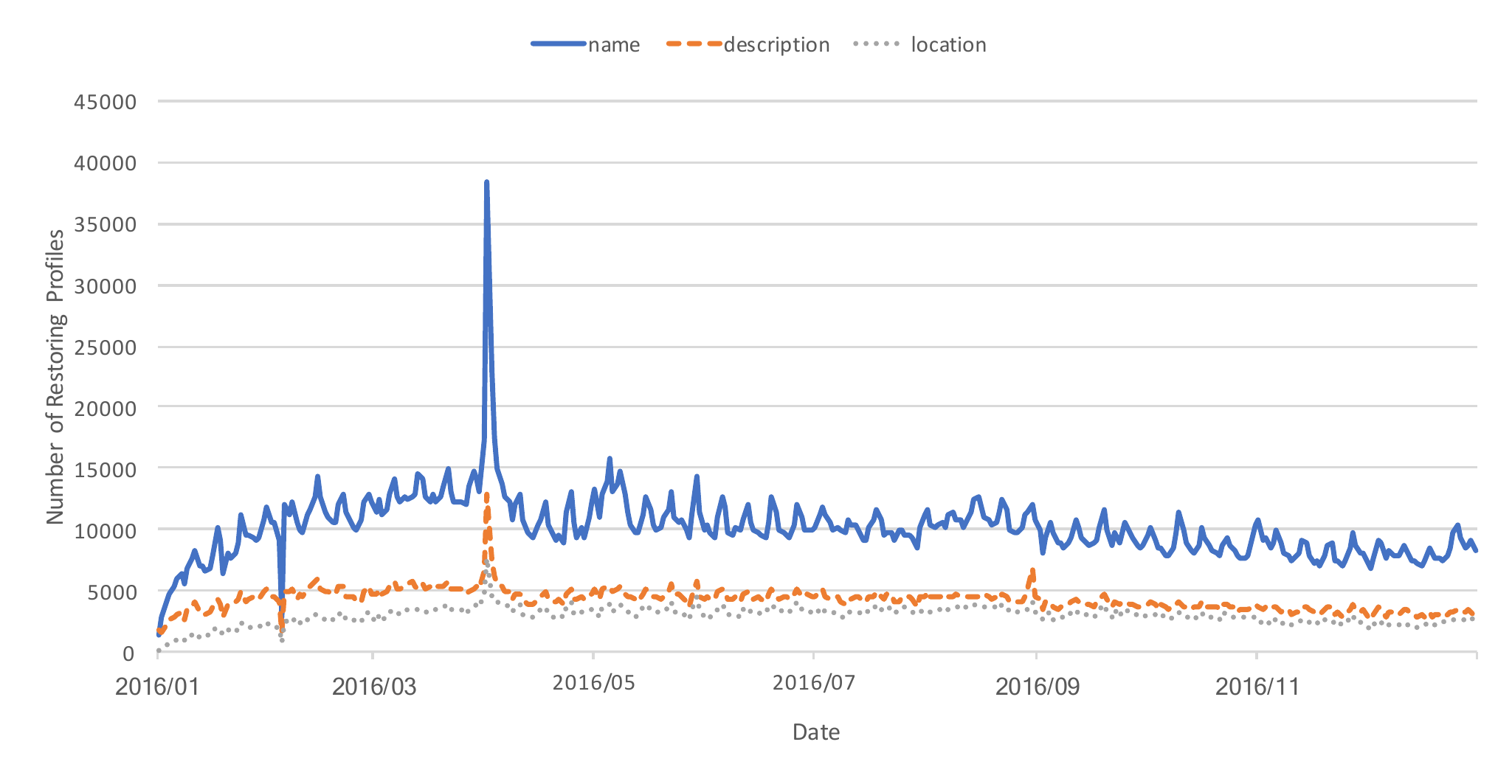}
    \label{fig:restore-profile-ja}
  }
  \hfil
  \subfloat[English (en)]{
    \includegraphics[width=1\linewidth]{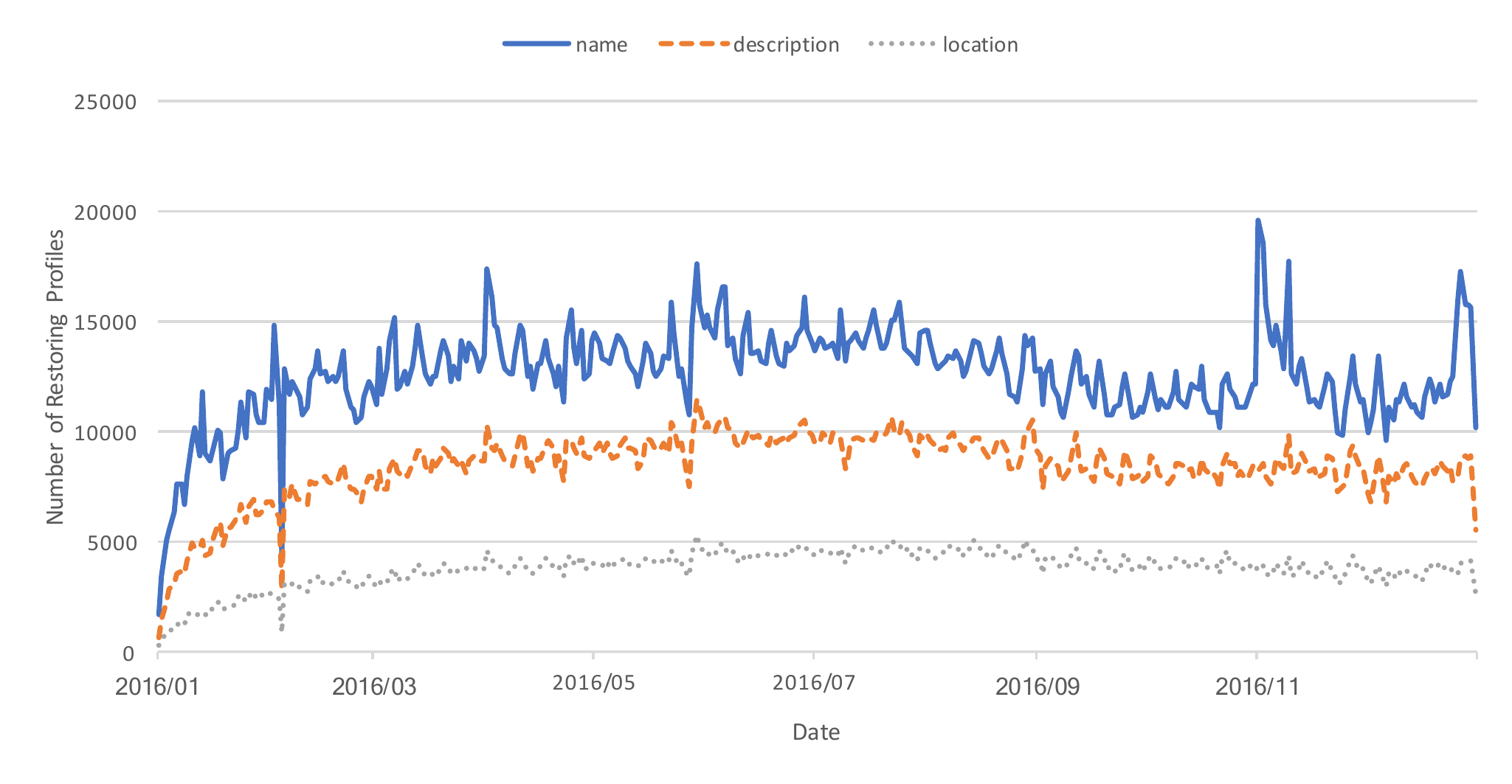}
    \label{fig:restore-profile-en}
  }
  \caption{Trend of restoring profile information: a sharp peak occurs on April 1 in the name of Japanese. This shows that a special name is used only on April 1.}
  \label{fig:restore-profile}
\end{figure}

For transient events like April Fools' Day, users may quickly restore their profile information.
We investigated when the users restored the profile information like A $\to$ B $\to$ A.
The result is shown in Fig.~\ref{fig:restore-profile}.
The vertical axis indicates the retweets whose profile information was restored on that day.

The name was the most restorable in both Japanese and English.
We found that a sharp peak occurs on April 1 in the name of Japanese.
This shows that a special name is used only on April 1.
In other words, it seems that the April Fools' Day influence was greater than a new life starting.
We also found that a peak occurs on November 1\footnote{Since the data is counted in JST, it corresponds to October 31 in the United States.} in the names in English.
This may be the effect of Halloween.
The second peak was November 9\footnote{It corresponds to November 8 in the United States.} in English.
This day is consistent with the date of the presidential election in the United States.

\subsection{RQ2: Keywords Inserted When Changing Profiles}

In the previous section, we only investigated the date when profile information was changed.
In this section, we investigate what kind of keywords are inserted when changing profile information.
In order to extract the inserted keywords when changing profile information,
we make character strings before and after the profile change, and extract the difference of the strings as a keyword.
In this analysis, we only report changes in the name.
A universal noun and a place name appeared in the description and the location but this was not a characteristic result.

\subsubsection{Japanese Retweets and Users}

Keyword extraction was executed as follows.
First, we use Python class ``difflib.SequenceMatcher'' to extract common strings of character strings before and after the profile change.
Next, we divide the string after the profile change by the common strings.
Each divided character string is an inserted keyword candidate.
We regard candidates with lengths greater than four characters as inserted keywords.
For example, the inserted keyword is ``固定ツイ''
when the string before the profile change is ``しましま@低浮上'' and the string after the profile change is ``しましま@固定ツイ''.

In TABLE~\ref{tb:name-ja}, we show the top 20 keywords inserted when changing the name in Japanese.
Characteristic keywords are underlined.
The keywords in the table were sorted by the number of occurrences (retweets), and the number of unique users was also shown.
English translation by the authors is added to the table.
Many users are inserting ``低浮上'' (Tei-fujo; Activity goes down) and ``固定ツイ'' (Kotei-tsui; Pinned tweet).
These words are often used by young people, especially students.
Students cannot make time to tweet when they are busy, such as when they are preparing for an examination.
The users use ``低浮上'' to indicate to friends that their Twitter engagement and activities decrease over time. 
The users also describe recent status to the pinned tweet\footnote{This is the tweet pinned one of user's tweets to the top of him/her page.}, and appeal contents to friends that cannot be described by the name alone.
In such a case ``固定ツイ'' is often inserted.

\begin{table}[tp]
  \centering
  \caption{Top 20 keywords inserted when changing the name in Japanese.}
  \begin{tabular}{ l r r | l }
\hline
inserted keyword & retweets & users & English translation \\
\hline \hline
\underline{@低浮上} & 24,659 & 20,903 & Activity goes down \\
ありがとう & 12,088 & 10,820 & Thank you \\
お疲れ様でした & 9,745 & 8,150 & Good job \\
\underline{固定ツイ} & 8,340 & 6,990 & Pinned tweet \\
\underline{は低浮上} & 6,511 & 5,734 & Activity goes down \\
当選祈願 & 5,175 & 4,653 & Prize winning prayer \\
\underline{（低浮上）} & 5,159 & 4,710 & Activity goes down \\
テスト期間 & 5,070 & 4,573 & Examination period \\
お疲れ様でした！ & 4,951 & 4,182 & Good job! \\
\underline{@固定ツイ} & 3,250 & 2,767 & Pinned tweet \\
\underline{[반동결]} & 3,193 & 2,675 & Activity goes down \\
（アローラのすがた） & 2,762 & 2,671 & Alola Form (Pokémon) \\
\underline{固定ツイ見てね} & 2,640 & 2,415 & Show pinned tweet \\
楽しかった & 2,488 & 2,342 & I was fun \\
おじさん & 2,342 & 2,059 & Daddy \\
お疲れ様 & 2,332 & 2,116 & Good job \\
\underline{【低浮上】} & 2,311 & 2,045 & Activity goes down \\
しんどい & 2,311 & 2,173 & I am tired \\
†┏┛墓┗┓† & 2,201 & 1,955 & Grave \\
\underline{低浮上気味} & 2,125 & 1,986 & Activity goes down \\
\hline
  \end{tabular}
  \label{tb:name-ja}
\end{table}

We also observed the results for each month.
The top five keywords were largely fixed,
but time specific keywords such as ``ポケモンGO'' (Pokémon Go), ``の名は。'' (Substring of movie title ``Your Name.''), ``ハロウィン'' (Halloween) and ``クリスマス'' (Christmas) appear in the top 20 keywords.
Users add both seasonal event names and their own status to their names.

\subsubsection{English Retweets and Users}

Keyword extraction was executed as follows.
First, strings of character strings before and after the profile change are divided into words with space characters.
Let an inserted keyword candidate be the word that appeared in the string after the profile change but did not appear in the string before the profile change.
We regard candidates as those including alphabet characters and lengths greater than four characters as inserted keywords.
For example, the inserted keyword is ``sunny''
when the string before the profile change is ``It's raining today.'' and the string after the profile change is ``It is sunny today.''.

In TABLE ~\ref{tb:name-en}, we show the top 20 keywords inserted when changing the name in English.
Characteristic keywords are underlined.
The keywords in the table were sorted by the number of occurrences (retweets), and the number of unique users was also shown.
Many users insert keywords related to birthdays (e.g., BIRTHDAY, birthday and BDAY).
This is a trend not observed at all in Japanese.

\begin{table}[tp]
  \centering
  \caption{Top 20 keywords inserted when changing the name in English.}
  \begin{tabular}{ l r r }
\hline
inserted keyword & retweets & users  \\
\hline \hline
loves & 63,865 & 23,086 \\
\underline{HAPPY} & 33,267 & 20,276 \\
\underline{happy} & 26,037 & 15,935 \\
\underline{BDAY} & 21,677 & 13,866 \\
\underline{bday} & 19,069 & 13,793 \\
\underline{BIRTHDAY} & 16,351 & 13,742 \\
TODAY & 16,200 & 12,316 \\
\underline{birthday} & 15,849 & 14,239 \\
girl & 14,885 & 12,877 \\
days & 14,465 & 10,199 \\
pinned & 13,221 & 7,251 \\
misses & 12,112 & 7,588 \\
\underline{Happy} & 11,248 & 8,290 \\
baby & 10,796 & 7,962 \\
love & 10,289 & 8,457 \\
spooky & 10,148 & 8,493 \\
7/27 & 9,390 & 5,030 \\
PLEASE & 9,246 & 5,413 \\
TOMORROW & 9,232 & 7,424 \\
stan & 9,146 & 6,263 \\
\hline
  \end{tabular}
  \label{tb:name-en}
\end{table}

We also observed the results for each month.
The top five keywords were largely fixed,
but time specific keywords such as ``spooky'', ``Christmas'' and ``Merry'' appear in the top 20 keywords.
September to November 2016 was the presidential election season in the United States.
During this season, keywords related to elections such as ``Deplorable'', ``VOTE'' and ``Trump'' appeared in the top results.

\section{Conclusion}

We focused on user's profile information changes and analyzed the timing and reasons for change on Twitter.
The results indicated that the peak of profile information change occurs in April among Japanese users, but there was no such trend observed for English users throughout the year.
Our analysis also showed that English users most frequently change their names on their birthdays, while Japanese users change their names as their Twitter engagement and activities decrease over time.

\bibliographystyle{IEEEtran}
\bibliography{IEEEabrv,references}

\end{document}